\documentclass[preprintnumbers,amsmath,amssymb,superscriptaddress]{revtex4-1}
\usepackage{bm}
\usepackage{graphicx}
\usepackage{tikz}
\usepackage{chemarrow}
\usepackage{multirow}
\begin{document}

%----------------------------------------------------------------------------------------------
\title{Homochiral oligopeptides by chiral
amplification: Interpretation of experimental data with a
copolymerization model.}
%----------------------------------------------------------------------------------------------

\author{Celia Blanco}
\email{blancodtc@cab.inta-csic.es} \affiliation{Centro de
Astrobiolog\'{\i}a (CSIC-INTA), Carretera Ajalvir Kil\'{o}metro 4,
28850 Torrej\'{o}n de Ardoz, Madrid, Spain}
\author{David Hochberg}
\email{hochbergd@cab.inta-csic.es} \affiliation{Centro de
Astrobiolog\'{\i}a (CSIC-INTA), Carretera Ajalvir Kil\'{o}metro 4,
28850 Torrej\'{o}n de Ardoz, Madrid, Spain}

\begin{abstract}
We present a differential rate equation model of chiral
polymerization based on a simple copolymerization scheme in which
the enantiomers are added to, or removed from, the homochiral or
heterochiral chains (reversible stepwise isodesmic growth or
dissociation). The model is set up for closed systems and takes into
account the corresponding thermodynamic constraints implied by the
reversible monomer attachments, while obeying a constant mass
constraint. In its simplest form, the model depends on a single
variable rate constant, the maximum chain length N, and the initial
concentrations. We have fit the model to the experimental data from
the Rehovot group on lattice-controlled chiral amplification of
oligopeptides. We find in all the chemical systems employed except
for one, that the model fits the measured relative abundances of the
oligopetides with higher degrees of correlation than from a purely
random polymerization process.
\end{abstract}

\pacs{05.40.Ca, 11.30.Qc, 87.15.B-}
\date{\today}

\maketitle

\section{Introduction}

In the transition from prebiotic racemic chemistry to chiral biology
one scenario suggests that homochiral peptides must have appeared
before the appearance of the primeval enzymes
\cite{Joyce,Avetisov1996}. While several stochastic synthetic routes
for mirror symmetry breaking that convert racemates into
nonracemates have been described \cite{KondeAsak,Cintas}, the
generation of long bio-like polymers \cite{Joyce} made up of
repeating units of the same handedness requires elaboration of new
synthetic routes. Polymerization reactions of racemic mixtures of
monomers in solution are typically expected to yield polymers
composed of random sequences of the left- and right-handed repeat
units following a binomial or Bernoulli distribution. Thus the
probability for obtaining oligomers with homochiral sequence becomes
negligible with increasing length \cite{Joyce}.

Recent investigations have proposed that $N$-carboxyanhydride
(NCA)\cite{Huber,Leman} and thioester derivatives
\cite{Huber,deDuve} of amino acids might have operated as relevant
precursors for the formation of the early peptides \cite{Pascal}.
Results on the polymerization of NCA monomers in organic
solvents,\cite{Lundberg,Idelson,Akaike,Blair1,Blair2,Krich} in
water\cite{Ehler,Hill,Brack} and in the solid state
\cite{Kanaz1,Kanaz2} have been published. Luisi and
coworkers\cite{Hitz2001,Hitz2002,HL2003,BHL} have reported the
polymerization of racemic $\alpha$-amino acids in solution which
yields small amounts of oligopeptides of homochiral sequence whose
abundances with respect to the heterochiral chains exhibit a slight
departure from the binomial distribution.

This problem of the random distribution can be overcome by catalyzed
polymerization of amphiphilic amino acids, in racemic and nonracemic
forms, which self-assemble into two-dimensional ordered crystallites
at the air-water interface \cite{Lahav2002b,Lahav2003b}.  Based on a
process involving self-assembly followed by lattice controlled
polymerization, Lahav and coworkers recently proposed a general
scenario for the generation of homochiral oligopeptides of a single
handedness from non-racemic mixtures of activated alpha amino acids
\cite{Lahav2002b,Lahav2003b}. Initial non-racemic mixtures undergo a
phase separation by self-assembly into a 2D racemic crystalline
phase and a separate enantiomorphous 2D phase of the enantiomer in
excess. Each of these crystalline phases has markedly different
chemical properties, thus yielding products that differ in the
composition of the oligomers. So, polymerization within the
enantiomorphous crystalline phase yields homochiral oligopeptides of
one handedness whereas the reaction controlled by the racemic
crystallites yields racemic mixtures and heterochiral products. The
combination of the two routes leads to an overall chiral
amplification process.

In this paper, we are interested in the lattice-controlled
polymerization reactions proposed by those authors. It is important
to clarify at the outset what specific aspect of the overall
experimental mechanism we want model here and the way we aim to do
so. The proposed experimental scheme starts from an initial excess,
say $S>R$ of monomers which undergoes an initial self-assembly
process into two types of two-dimensional crystallites at the
air/water interface. Once formed, each one of these two crystal
phases participates in the control of a subsequent type of
polymerization. Thus, the racemic crystallites polymerize racemic
mixtures of oligomers and the heterochiral products, whereas the
other pure enantiomorphous crystallite controls the polymerization
of  the isotactic chains, these are formed from the monomer in
excess ($S$, in this example). However, the details of the
polymerizations depend in a complicated way upon the specific
\textit{packing arrangements} of the crystal monomers and the
possible \textit{reaction pathways} taken within each crystallite
phase. The authors of the experiments state that the connection
between the monomer packing arrangements in the crystallites and the
resultant composition of the various diastereoisomeric products is
``not straighforward" \cite{Lahav2003b}. We therefore opt for a
simple model for interpreting their data. With this objective in
mind, we present a copolymerization model for the interpretation of
the experimental data. The model may be termed \textit{effective} in
the following sense: it presupposes or takes as given the
\textit{prior} formation of the self-assembled 2D crystallites at
the air-water interface and is concerned exclusively with the
subsequent polymerization reactions. Thus the complicated
microscopic details referring to the monomer packing arrangements
and reaction pathways within the crystallite self-assemblies are
treated implicitly with our rate constants. Our copolymerization
reaction rates can satisfactorily account for the different chemical
properties of the two crystalline phases (racemic 2D crystallites
and pure enantiomorphous 2D crystallites) that lead to the formation
of racemic mixtures, heterochiral products and isotactic
oligopeptides. We contrast the fits from our model with those
assuming a purely random process that obeys a binomial distribution.
The final justification for considering such an effective model
rests on its ability to yield good fits to the data. The goodness of
the fits obtained below demonstrates that the experimental data can
be fit convincingly as if the simple scheme depicted pictorially in
Fig. \ref{scheme} were the sole mechanism leading to the observed
relative abundances. This then gives additional meaning to to term
``effective", and in the operational sense.

%Footnotes

%Please use \dag to cite the ESI in the main text of the article.
%If you article does not have ESI please remove the \dag symbol from the title and the above footnotetext.

\footnotetext{\textit{$^{a}$~Centro de Astrobiolog\'{\i}a
(CSIC-INTA), Ctra. Ajalvir Km. 4, 28850 Torrej\'{o}n de Ardoz,
Madrid, Spain. E-mail: blancodtc@cab.inta-csic.es,
hochbergd@cab.inta-csic.es.}}

%additional addresses can be cited as above using the lower-case letters, c, d, e... If all authors are from the same address, no letter is required

%----------------------------------------------------------------
\section{\label{sec:model} The copolymerization model}
%---------------------------------------------------------------

\begin{figure}
\centering
  \includegraphics[width=0.45\textwidth]{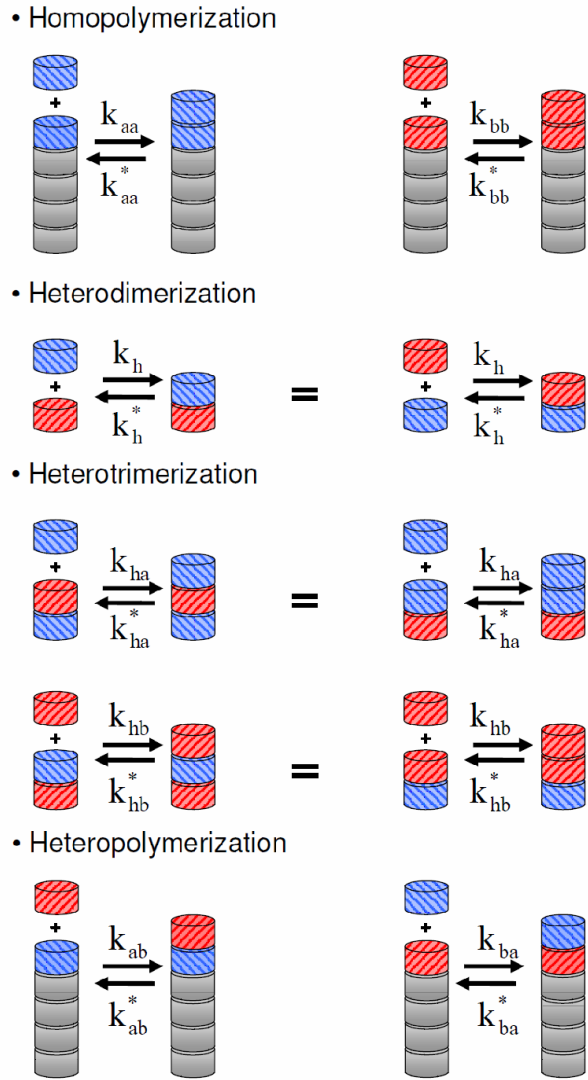}
  \caption{The copolymerization model. The (R)-chiral (red) and (S)-chiral (blue) monomers
  reversibly associate into the growing homochiral (top) or heterochiral (bottom) copolymer chains.
  Because the system is \textit{closed}, both the heterodimer (second line) and hetero-trimer
  (third and fourth lines) reactions must be treated separately to avoid
  double counting and thus ensure that the total system mass is conserved in a closed system (see text for an explanation).}
  \label{scheme}
\end{figure}

Our starting point is a simple model for the copolymerization of two
chemically distinct monomers displaying a wide variety of product
sequence compositions. The model we introduce and study here is an
appropriately modified and extended version of the one considered a
few years ago by Wattis and Coveney \cite{WC2}.

The main important differences compared to prior and related models
are that we (1) consider polymerization in \textit{closed} systems
\cite{BH}--, so that no matter flow is permitted with an external
environment-- and (2) we allow for reversible monomer association
steps. We  also correctly include the formation (and dissociation)
of the heterodimer \cite{BH}. It turns out this must be treated on a
separate basis in order to avoid double counting, which if left
unchecked, would lead to a violation in the constant mass
constraint. Once the heterodimer is treated correctly, this implies
that the hetero-trimer must also be treated separately. Beyond this,
the remainder of the hetero-oligomers can be treated in a uniform
way.

First, we introduce the notation to be used. Polymers are classified
by three quantities: the number of A monomers of which it is
composed (subscript $r$), the number of B monomers which it contains
(subscript $s$) and the final or terminal monomer in the chain,
denoted by a superscript. In this scheme, the monomers are denoted
by $A=C^A_{1,0}$ and $B=C^B_{0,1}$ ; pure homopolymers are denoted
by $C^A_{r,0}$ and $C^B_{0,s}$ ; all copolymer chains $C^A_{r,s}$ or
$C^B_{r,s}$ with $r, s \geq 1$ are heteropolymers. Note also that
chains of the form $C^A_{0,s}$ and $C^B_{r,0}$ are forbidden. The
corresponding time-dependent concentrations are denoted by lower
case variables: e.g.,  $c^A_{r,s}(t)$ and $c^B_{r,s}(t)$. The model
is then defined by the following reactions, in which equilibrium is
maintained between the finite monomer pool and the ensemble of
copolymers:
\begin{eqnarray}\label{copolyscheme}
C_{r,s}^A + A &\autorightleftharpoons{$k_{aa}$}{$k_{aa}^{*}$}&
C_{r+1,s}^A, \\
C_{r,s}^A + B &\autorightleftharpoons{$k_{ab}$}{$k_{ab}^{*}$}&
C_{r,s+1}^B,\\
C_{r,s}^B + A &\autorightleftharpoons{$k_{ba}$}{$k_{ba}^{*}$}&
C_{r+1,s}^A,\\
C_{r,s}^B + B &\autorightleftharpoons{$k_{bb}$}{$k_{bb}^{*}$}&
C_{r,s+1}^B.
\end{eqnarray}
This model can accommodate any two chemically distinct monomers. For
the purpose of this paper, we consider the case when $A = R$ and $B
= S$ are two enantiomers.

The overall basic scheme must be broken down into several special
subcases, especially important so as to avoid undesired double
counting of the heterodimer and heterotrimer reactions, see  Fig.
\ref{scheme}. Once we treat these special cases, we then pass to the
corresponding set of rate equations for the concentrations.

The formation of chirally pure polymer chains denoted by $c^A_{n,0}$
and $c^B_{0,n}$, for $1\leq n \leq N-1$ is described by the
homo-polymerization reactions:

\begin{eqnarray}\label{homo}
C_{n,0}^{A} + C_{1,0}^{A}
\autorightleftharpoons{$k_{aa}$}{$k_{aa}^{*}$} C_{n+1,0}^{A}&&\qquad
C_{0,n}^{B} + C_{0,1}^{B}
\autorightleftharpoons{$k_{bb}$}{$k_{bb}^{*}$}
C_{0,n+1}^{B}.\nonumber\\
\end{eqnarray}
$N$ is the maximum chain length permitted. In our recently reported
work \cite{BH}, we considered that once a monomer has been added to
a homopolymer of the opposite chirality (that is, "the wrong"
monomer), the polymer is inhibited and further growth is halted.
This polymer could not directly react anymore and could only lose
its wrong terminal monomer through the inverse reaction. In the
present model, we assume such a chain can continue to grow by adding
monomers of both configurations. So, for $2\leq n\leq N-1$, the
hetero-polymerization or inhibition reactions are as follows:
\begin{eqnarray}\label{hetero1}
C_{n,0}^{A} + C_{0,1}^{B}
\autorightleftharpoons{$k_{ab}$}{$k_{ab}^{*}$} C_{n,1}^{B}&&\qquad
C_{0,n}^{B} + C_{1,0}^{A}
\autorightleftharpoons{$k_{ba}$}{$k_{ba}^{*}$}
C_{1,n}^{A}.\nonumber\\
\end{eqnarray}
For both homo- and hetero-polymerization reactions, represented by
Eq. \ref{homo}-\ref{hetero1}, the upper limits specified for $n$
ensure that the \textit{maximum} length for all oligomers produced
(or consumed) by these reaction sets, both the homo- and
heterochiral ones, is never greater than $N$. In the remainder of
this paper we will consider here the natural and chiral symmetric
reaction rate assignments $k_{aa}=k_{bb}$, $k_{ab}=k_{ba}$ and
likewise for the inverse rates, $k^{*}_{aa}=k^{*}_{bb}$ and
$k^{*}_{ab}=k^{*}_{ba}$, reducing the number of independent rate
constants to four.

Even if we have the information about the composition, we can only
know the chirality of the last monomer attached to the chain, we
have no information regarding the specific \textit{sequence}. This
implies that the following two reactions are indistinguishable:
\begin{equation}
C_{1,0}^{A} + C_{0,1}^{B}
\autorightleftharpoons{$k_{ab}$}{$k_{ab}^{*}$} C_{1,1}^{B}\qquad
C_{0,1}^{B} + C_{1,0}^{A}
\autorightleftharpoons{$k_{ba}$}{$k_{ba}^{*}$} C_{1,1}^{A}.
\nonumber
\end{equation}
Thus for all practical purposes, $C_{1,1}^{A} \equiv C_{1,1}^{B}$,
and this suggests using the following notation: $C_{1,1} \equiv
C_{1,1}^{A} \equiv C_{1,1}^{B}$, and to define a unique direct
constant rate: $k_{h}=\frac{k_{ab}+k_{ba}}{2}$, and an inverse one
$k_{h}^{*}=\frac{k_{ab}^{*}+k_{ba}^{*}}{2}$. Note that if
$k_{ab}=k_{ba}$, then $k_{h}=k_{ab}=k_{ba}$. Due to these
characteristics, we will treat the heterodimer in a different way
compared with the other hetero-polymers. The reaction of the
heterodimer formation is therefore:
\begin{equation}\label{heterodimer}
C_{1,0}^{A} + C_{0,1}^{B}
\autorightleftharpoons{$k_{h}$}{$k_{h}^{*}$} C_{1,1}.
\end{equation}
As before, the reactives and products in Eq. (\ref{heterodimer}) are
the same, so the differences in the free energy between initial and
final states should be the same in all the reactions in these
equations, implying the following thermodynamic constraint on the
reaction rates:

\begin{equation}\label{constraint2}
\frac{k_{ab}}{k_{ab}^{*}}=\frac{k_{ba}}{k_{ba}^{*}}.
\end{equation}
If the heterodimer formation were not to be treated in the separate
way as we have done, and were to be included, e.g. in
Eq.(\ref{hetero1}) by merely changing the lower limits for $n$
($2\leq n\leq N-1$) by $1\leq n\leq N-1$, we would be making the
mistake of double counting it. The same occurs for the
heteropolymers formed from the addition of a monomer to a
heterodimer. The two reactions of each pair of the following
equations are also indistinguishable:
\begin{equation}
C_{1,1} + C_{1,0}^{A} \autorightleftharpoons{$k_{aa}$}{$k_{aa}^{*}$}
C_{2,1}^{A}\qquad\ C_{1,1} + C_{1,0}^{A}
\autorightleftharpoons{$k_{ba}$}{$k_{ba}^{*}$} C_{2,1}^{A} \nonumber
\end{equation}

\begin{equation}
C_{1,1} + C_{0,1}^{B} \autorightleftharpoons{$k_{bb}$}{$k_{bb}^{*}$}
C_{1,2}^{B}\qquad C_{1,1} + C_{0,1}^{B}
\autorightleftharpoons{$k_{ab}$}{$k_{ab}^{*}$} C_{1,2}^{B}. \nonumber\\
\end{equation}

Again, it is convenient to define the following direct reaction
rates for these steps, $k_{ha}=\frac{k_{aa}+k_{ba}}{2}$,
$k_{hb}=\frac{k_{bb}+k_{ab}}{2}$ and inverse
$k_{ha}^{*}=\frac{k_{aa}^{*}+k_{ba}^{*}}{2}$,
$k_{hb}^{*}=\frac{k_{bb}^{*}+k_{ab}^{*}}{2}$ . Note that if
$k_{aa}=k_{bb}$ and $k_{ab}=k_{ba}$, then $k_{ha}=k_{hb}$, and if
$k_{aa}^{*}=k_{bb}^{*}$ and $k_{ab}^{*}=k_{ba}^{*}$, then
$k_{ha}^{*}=k_{hb}^{*}$. The reactions to consider are then:
\begin{equation}\label{trimer}
C_{1,1} + C_{1,0}^{A} \autorightleftharpoons{$k_{ha}$}{$k_{ha}^{*}$}
C_{2,1}^A\qquad C_{1,1} + C_{0,1}^{B}
\autorightleftharpoons{$k_{hb}$}{$k_{hb}^{*}$} C_{1,2}^B.
\end{equation}
As we have already remarked, in our model, as in the original one
for open systems \cite{WC2}, the polymeric chains that have taken up
the "wrong" chirality monomer can continue to grow. Thus, we we
allow for the further growth of these chains by adding monomers of
either chirality. This kind of polymerization reaction for $2\leq
n\leq N-2$ is given by:
\begin{eqnarray}\label{hetero2s}
C_{1,n}^{A} + C_{1,0}^{A}
\autorightleftharpoons{$k_{aa}$}{$k_{aa}^{*}$} C_{2,n}^{A}\qquad
C_{n,1}^{B} + C_{0,1}^{B}
\autorightleftharpoons{$k_{bb}$}{$k_{bb}^{*}$} C_{n,2}^{B}
\end{eqnarray}
\begin{eqnarray}\label{hetero2o}
C_{1,n}^{A} + C_{0,1}^{B}
\autorightleftharpoons{$k_{ab}$}{$k_{ab}^{*}$} C_{1,n+1}^{B}\qquad
C_{n,1}^{B} + C_{1,0}^{A}
\autorightleftharpoons{$k_{ba}$}{$k_{ba}^{*}$} C_{n+1,1}^{A}.
\end{eqnarray}\\

And for $2\leq r\leq N-2$ , $1\leq s\leq N-1-r$:

\begin{eqnarray}\label{hetero3s}
C_{r,s}^{A} + C_{1,0}^{A}
\autorightleftharpoons{$k_{aa}$}{$k_{aa}^{*}$} C_{r+1,s}^{A}\qquad
C_{r,s}^{B} + C_{0,1}^{B}
\autorightleftharpoons{$k_{bb}$}{$k_{bb}^{*}$} C_{r,s+1}^{B}
\end{eqnarray}

\begin{eqnarray}\label{hetero3o}
C_{r,s}^{A} + C_{0,1}^{B}
\autorightleftharpoons{$k_{ab}$}{$k_{ab}^{*}$} C_{r,s+1}^{B}\qquad
C_{r,s}^{B} + C_{1,0}^{A}
\autorightleftharpoons{$k_{ba}$}{$k_{ba}^{*}$} C_{r+1,s}^{B}.
\end{eqnarray}
Note that in the elementary reaction steps, in the rate constants,
and in the corresponding differential rate equations (see below),
the left-right symmetry of the model is manifest, that is, possesses
a discrete $Z_2$ symmetry. This symmetry can be broken spontaneously
by the dynamical solutions of the differential rate equations, thus
this model is apt for studying spontaneous mirror symmetry breaking.

By lifting the $Z_2$ degeneracy in the reaction rates, e.g.,
allowing for $k_{aa} \neq k_{bb}$ and thus leading to more
independent rate constants for describing the reaction set, we could
study the influence of explicit chiral bias in the model. As this is
not the aim of this work, we will not consider it here.

We next write down the differential rate equations corresponding to
this reaction network.  We employ the rate-equation theory as in
chemical kinetics. We begin with the rate equations for the chiral
monomers:

\begin{eqnarray}\label{dA}
\frac{d c_{1,0}^{A}}{dt}&=& - k_{aa}c_{1,0}^{A}\Big(2c_{1,0}^{A}
+\sum_{n=2}^{N-1}c_{n,0}^{A}+\sum_{n=2}^{N-2}c_{1,n}^{A}
+\sum_{r=2}^{N-2}\sum_{s=1}^{N-1-r}c_{r,s}^{A}\Big)-k_{ba}c_{1,0}^{A}\Big(\sum_{n=2}^{N-1}c_{0,n}^{B}+\sum_{n=2}^{N-2}c_{n,1}^{B}
+\sum_{s=2}^{N-2}\sum_{r=1}^{N-1-s}c_{r,s}^{B}\Big)\nonumber\\
&-&k_{h}c_{1,0}^{A}c_{0,1}^{B}-k_{ha}c_{1,0}^{A}c_{1,1}+k_{aa}^{*}\Big(2c_{2,0}^{A}+\sum_{n=3}^{N}c_{n,0}^{A}+\sum_{n=2}^{N-2}c_{2,n}^{A}+\sum_{r=3}^{N-1}\sum_{s=1}^{N-r}c_{r,s}^{A}\Big)\nonumber\\
&+&k_{ba}^{*}\Big(\sum_{n=2}^{N-1}c_{1,n}^{A}+\sum_{n=2}^{N-2}c_{2,n}^{A}+\sum_{r=3}^{N-1}\sum_{s=1}^{N-r}c_{r,s}^{A}\Big)
+k_{h}^{*}c_{1,1}+k_{ha}^{*}c_{2,1}^{A}\\
\nonumber\end{eqnarray}

\begin{eqnarray}\label{dB}
\frac{d c_{0,1}^{B} }{dt}&=&-k_{bb}c_{0,1}^{B}\Big(2c_{0,1}^{B}
+\sum_{n=2}^{N-1}c_{0,n}^{B}+\sum_{n=2}^{N-2}c_{n,1}^{B}
+\sum_{s=2}^{N-2}\sum_{r=1}^{N-1-s}c_{r,s}^{B}\Big)-k_{ab}c_{0,1}^{B}\Big(\sum_{n=2}^{N-1}c_{n,0}^{A}+\sum_{n=2}^{N-2}c_{1,n}^{A}
+\sum_{r=2}^{N-2}\sum_{s=1}^{N-1-r}c_{r,s}^{A}\Big)\nonumber\\
&-&k_{h}c_{1,0}^{A}c_{0,1}^{B}-k_{hb}c_{0,1}^{B}c_{1,1}+k_{bb}^{*}\Big(2c_{0,2}^{B}+\sum_{n=3}^{N}c_{0,n}^{B}+\sum_{n=2}^{N-2}c_{n,2}^{B}+\sum_{s=3}^{N-1}\sum_{r=1}^{N-s}c_{r,s}^{B}\Big)\nonumber\\
&+&k_{ab}^{*}\Big(\sum_{n=2}^{N-1}c_{n,1}^{B}+\sum_{n=2}^{N-2}c_{n,2}^{B}+\sum_{s=3}^{N-1}\sum_{r=1}^{N-s}c_{r,s}^{B}\Big)+k_{h}^{*}c_{1,1}+k_{hb}^{*}c_{1,2}^{B}\\
\nonumber\end{eqnarray}\\

The equations describing the concentration of the homopolymers, for
$2\leq n \leq N-1$:

\begin{eqnarray}\label{dhomo}
\frac{d
c_{n,0}^{A}}{dt}&=&k_{aa}c_{1,0}^{A}\Big(c_{n-1,0}^{A}-c_{n,0}^{A}\Big)-k_{ab}c_{n,0}^{A}c_{0,1}^{B}+k_{aa}^{*}\Big(c_{n+1,0}^{A}-c_{n,0}^{A}\Big)+k_{ab}^{*}c_{n,1}^{B}\\
\frac{d
c_{0,n}^{B}}{dt}&=&k_{bb}c_{0,1}^{B}\Big(c_{0,n-1}^{B}-c_{0,n}^{B}\Big)-k_{ba}c_{0,n}^{B}c_{1,0}^{A}+k_{bb}^{*}\Big(c_{0,n+1}^{B}-c_{0,n}^{B}\Big)+k_{ba}^{*}c_{1,n}^{A}\\
\nonumber\end{eqnarray}
It is necessary to treat the kinetic equations of the maximum length
homopolymers $N$ individually. Since these do not elongate further,
they can not directly react, and can not be the product of an
inverse reaction involving a longer chain:
\begin{eqnarray}\label{dhomomax}
\frac{d c_{N,0}^{A} }{dt}&=&k_{aa}c_{1,0}^{A}c_{N-1,0}^{A}-k_{aa}^{*}c_{N,0}^{A}\\
\frac{d c_{0,N}^{B}}{dt}&=&k_{bb}c_{0,1}^{B}c_{0,N-1}^{B}-k_{bb}^{*}c_{0,N}^{B}\\
\nonumber\end{eqnarray}
The differential equations describing the concentration of each type
of heteropolymer (included the heterodimer), for $2\leq n\leq N-2$ :

\begin{eqnarray}\label{dhetero1}
\frac{d c_{1,1}
}{dt}&=&k_{h}c_{1,0}^{A}c_{0,1}^{B}-k_{ha}c_{1,1}c_{1,0}^{A}
-k_{hb}c_{1,1}c_{0,1}^{B}-k_{h}^{*}c_{1,1}+k_{ha}^{*}c_{2,1}^A+k_{hb}^{*}c_{1,2}^B\\
\frac{d
c_{1,n}^{A}}{dt}&=&-k_{aa}c_{1,0}^{A}c_{1,n}^{A}-k_{ab}c_{0,1}^{B}c_{1,n}^{A}
+k_{ba}c_{0,n}^{B}c_{1,0}^{A}+k_{aa}^{*}c_{2,n}^{A}+k_{ab}^{*}c_{1,n+1}^{B}-k_{ba}^{*}c_{1,n}^{A}\\
\frac{d
c_{n,1}^{B}}{dt}&=&-k_{bb}c_{0,1}^{B}c_{n,1}^{B}-k_{ba}c_{1,0}^{A}c_{n,1}^{B}
+k_{ab}c_{n,0}^{A}c_{0,1}^{B}+k_{bb}^{*}c_{n,2}^{B}+k_{ba}^{*}c_{n+1,1}^{A}-k_{ab}^{*}c_{n,1}^{B}\\
\nonumber\end{eqnarray}
As before, it is useful to treat individually the maximum length
polymers $N$:

\begin{eqnarray}\label{dhetero1max}
\frac{d c_{1,N-1}^{A} }{dt}&=&k_{ba}c_{0,N-1}^{B}c_{1,0}^{A}-k_{ba}^{*}c_{1,N-1}^{A}\\
\frac{d c_{N-1,1}^{B}}{dt}&=&k_{ab}c_{N-1,0}^{A}c_{0,1}^{B}-k_{ab}^{*}c_{N-1,1}^{B}\\
\nonumber\end{eqnarray}

As was mentioned when describing the reaction network, each kind of
trimer $c_{2,1}^{A}$ and $c_{1,2}^{B}$ must have its own
differential equation in terms of $k_{ha}$, $k_{hb}$:

\begin{eqnarray}\label{dtrimer}
\frac{d
c_{2,1}^{A}}{dt}&=&-k_{aa}c_{1,0}^{A}c_{2,1}^{A}-k_{ab}c_{0,1}^{B}c_{2,1}^{A}
+k_{ha}c_{1,1}c_{1,0}^{A}+k_{aa}^{*}c_{3,1}^{A}+k_{ab}^{*}c_{2,2}^{B}-k_{ha}^{*}c_{2,1}^{A}\\
\frac{d
c_{1,2}^{B}}{dt}&=&-k_{bb}c_{0,1}^{B}c_{1,2}^{B}-k_{ba}c_{1,0}^{A}c_{1,2}^{B}
+k_{hb}c_{1,1}c_{0,1}^{B}+k_{bb}^{*}c_{1,3}^{B}+k_{ba}^{*}c_{2,2}^{A}-k_{hb}^{*}c_{1,2}^{B}\\
\nonumber\end{eqnarray}

For $2\leq n\leq N-3$ :

\begin{eqnarray}\label{dhetero2}
\frac{d
c_{2,n}^{A}}{dt}&=&k_{aa}c_{1,0}^{A}\Big(c_{1,n}^{A}-c_{2,n}^{A}\Big)-k_{ab}c_{0,1}^{B}c_{2,n}^{A}+k_{ba}c_{1,n}^{B}c_{1,0}^{A}
+k_{aa}^{*}\Big(c_{3,n}^{A}-c_{2,n}^{A}\Big)+k_{ab}^{*}c_{2,n+1}^{B}-k_{ba}^{*}c_{2,n}^{A}\\
\frac{d
c_{n,2}^{B}}{dt}&=&k_{bb}c_{0,1}^{B}\Big(c_{n,1}^{B}-c_{n,2}^{B}\Big)-k_{ba}c_{1,0}^{A}c_{n,2}^{B}+k_{ab}c_{n,1}^{A}c_{0,1}^{B}+k_{bb}^{*}\Big(c_{n,3}^{B}-c_{n,2}^{B}\Big)+k_{ba}^{*}c_{n+1,2}^{A}-k_{ab}^{*}c_{n,2}^{B}\\
\nonumber\end{eqnarray}

Once again, the equations corresponding to the maximum length
homopolymers $N$ are
\begin{eqnarray}\label{dhetero2max}
\frac{d
c_{2,N-2}^{A}}{dt}&=&k_{aa}c_{1,0}^{A}c_{1,N-2}^{A}+k_{ba}c_{1,N-2}^{B}c_{1,0}^{A}-k_{aa}^{*}c_{2,N-2}^{A}-k_{ba}^{*}c_{2,N-2}^{A}\\
\frac{d
c_{N-2,2}^{B}}{dt}&=&k_{bb}c_{0,1}^{B}c_{N-2,1}^{B}+k_{ab}c_{N-2,1}^{A}c_{0,1}^{B}-k_{bb}^{*}c_{N-2,2}^{B}-k_{ab}^{*}c_{N-2,2}^{B}\\
\nonumber\end{eqnarray}

For $3\leq r\leq N-2$ and $1\leq s\leq N-1-r$:

\begin{eqnarray}\label{dhetero3}
\frac{d
c_{r,s}^{A}}{dt}&=&k_{aa}c_{1,0}^{A}\Big(c_{r-1,s}^{A}-c_{r,s}^{A}\Big)-k_{ab}c_{0,1}^{B}c_{r,s}^{A}+k_{ba}c_{r-1,s}^{B}c_{1,0}^{A}+k_{aa}^{*}\Big(c_{r+1,s}^{A}-c_{r,s}^{A}\Big)
+k_{ab}^{*}c_{r,s+1}^{B}-k_{ba}^{*}c_{r,s}^{A}\\
\end{eqnarray}

For $3\leq s\leq N-2$ and $1\leq r\leq N-1-s$:

\begin{eqnarray}\label{dhetero4}
\frac{d
c_{r,s}^{B}}{dt}&=&k_{bb}c_{0,1}^{B}\Big(c_{r,s-1}^{B}-c_{r,s}^{B}\Big)-k_{ba}c_{1,0}^{A}c_{r,s}^{B}+k_{ab}c_{r,s-1}^{A}c_{0,1}^{B}+k_{bb}^{*}\Big(c_{r,s+1}^{B}-c_{r,s}^{B}\Big)+k_{ba}^{*}c_{r+1,s}^{A}-k_{ab}^{*}c_{r,s}^{B}\\
\nonumber\end{eqnarray}

For $3\leq n\leq N-1$:

\begin{eqnarray}\label{dhetero3maxA}
\frac{d
c_{n,N-n}^{A}}{dt}&=&k_{aa}c_{1,0}^{A}c_{n-1,N-n}^{A}+k_{ba}c_{n-1,N-n}^{B}c_{1,0}^{A}-k_{aa}^{*}c_{n,N-n}^{A}-k_{ba}^{*}c_{n,N-n}^{A}\\
\label{dhetero3maxB} \frac{d
c_{N-n,n}^{B}}{dt}&=&k_{bb}c_{0,1}^{B}c_{N-n,n-1}^{B}+k_{ab}c_{N-n,n-1}^{A}c_{0,1}^{B}-k_{bb}^{*}c_{N-n,n}^{B}-k_{ab}^{*}c_{N-n,n}^{B}\\
\nonumber\end{eqnarray}
As remarked earlier, the complete reaction scheme must satisfy mass
conservation in a closed system, implying that the mass variation
rate must be strictly zero:
\begin{eqnarray}\label{mass}
0&=&2\dot{c}_{1,1}+3(\dot{c}^A_{2,1}+\dot{c}^B_{1,2})+\sum_{n=1}^{N}n\big(\dot{c}_{n,0}^{A}+\dot{c}_{0,n}^{B}\big)+\sum_{n=2}^{N-1}(n+1)\big(\dot{c}_{1,n}^{A}+\dot{c}_{n,1}^{B}\big)\nonumber\\
&+&\sum_{n=2}^{N-2}(n+2)\big(\dot{c}_{2,n}^{A}+\dot{c}_{n,2}^{B}\big)+\sum_{r=3}^{N-1}\sum_{s=1}^{N-1}(r+s)\big(\dot{c}_{r,s}^{A}+\dot{c}_{r,s}^{B}\big),
\end{eqnarray}
where the overdot stands for the time-derivative. The compliance
with this constraint is an important and crucial check on the
consistency of the numerical integration of the full set of
differential equations Eqs. (\ref{dA}-\ref{dhetero3maxB}), which we
monitor and confirm in all the simulations presented below.
Analytically, this relation is satisfied by the rate equations.

As we see, there is one differential equation for each type of
monomer and one for the heterodimer. The homopolymer set requires
$\sum_{n=2}2(N-1)$ equations and the heteropolymer set a total of
$2(N-2)$ equations. The total number of kinetic differential
equations describing the whole system is $N(N+1)$, and is broken
down into the separate contributions as displayed in Table
\ref{equations}.
\begin{table*}
\centering \small
  \caption{\ Number of differential equations as a function of the maximum polymer length $N$}
  \label{equations}
  \renewcommand{\arraystretch}{1.5}
\begin{tabular*}{\textwidth}{  c | c | c | c |} \cline{2-2}\cline{4-4}
  & \textbf{Number of eqs} & & \multicolumn{1}{c|}{\textbf{Number of eqs}} \\
  \cline{1-1}\cline{2-2}\cline{3-3}\cline{4-4}
  \multicolumn{1}{|c|}{$c^A_{1,0}$}
  & 1
  & \multicolumn{1}{c|}{$c^B_{0,1}$}
  & \multicolumn{1}{c|}{1} \\
  \cline{1-1}\cline{2-2}\cline{3-3}\cline{4-4}
  \multicolumn{1}{|c|}{$c_{1,1}$}
  &1
  &
  &  \\
  \cline{1-1}\cline{2-2}\cline{3-3}\cline{4-4}
  \multicolumn{1}{|c|}{$c^A_{n,0}$, $(2 \leq n \leq N)$}
  & \multicolumn{1}{c|}{$\sum_{n=2}^{N}=N-1$}
  &\multicolumn{1}{c|}{$c^B_{0,n}$, $(2 \leq n \leq N)$}
  & \multicolumn{1}{c|}{$\sum_{n=2}^{N}=N-1$} \\
  \cline{1-1}\cline{2-2}\cline{3-3}\cline{4-4}
  \multicolumn{1}{|c|}{$c^A_{1,n}$, $(2 \leq n \leq N-1)$}
  & \multicolumn{1}{c|}{$\sum_{n=2}^{N-1}=N-2$}
  &\multicolumn{1}{c|}{$c^B_{n,1}$, $(2 \leq n \leq N-1)$}
  & \multicolumn{1}{c|}{$\sum_{n=2}^{N-1}=N-2$} \\
  \cline{1-1}\cline{2-2}\cline{3-3}\cline{4-4}
  \multicolumn{1}{|c|}{$c^A_{2,1}$}
  & 1
  & \multicolumn{1}{c|}{$c^B_{1,2}$}
  & \multicolumn{1}{c|}{1} \\
  \cline{1-1}\cline{2-2}\cline{3-3}\cline{4-4}
  \multicolumn{1}{|c|}{$c^A_{2,n}$, $(2 \leq n \leq N-2)$}
  & \multicolumn{1}{c|}{$\sum_{n=2}^{N-2}=N-3$}
  &\multicolumn{1}{c|}{$c^B_{n,2}$, $(2 \leq n \leq N-2)$}
  & \multicolumn{1}{c|}{$\sum_{n=2}^{N-2}=N-3$} \\
  \cline{1-1}\cline{2-2}\cline{3-3}\cline{4-4}
  \multicolumn{1}{|c|}{$c^A_{r,s}$}
  & \multicolumn{1}{c|}{\multirow{3}{*}{$\sum_{r=3}^{N-2}\sum_{s=1}^{N-1-r}=\frac{1}{2}(N^2-7N+12)$}}
  &\multicolumn{1}{c|}{$c^B_{r,s}$}
  & \multicolumn{1}{c|}{\multirow{3}{*}{$\sum_{r=1}^{N-1-s}\sum_{s=3}^{N-2}=\frac{1}{2}(N^2-7N+12)$}}
  \\ \multicolumn{1}{|c|}{\footnotesize{$(3 \leq r \leq N-2)$}}
  & & \multicolumn{1}{c|}{\footnotesize{$(3 \leq s \leq N-2)$}} & \multicolumn{1}{c|}{} \\
  \multicolumn{1}{|c|}{\footnotesize{$(1 \leq s \leq
  N-1-r)$}}
  & & \multicolumn{1}{c|}{\footnotesize{$(1 \leq r \leq
  N-1-s)$}}& \multicolumn{1}{c|}{} \\
  \cline{1-1}\cline{2-2}\cline{3-3}\cline{4-4}
  \multicolumn{1}{|c|}{$c^A_{n,N-n}$, $(3 \leq n \leq N-1)$}
  & \multicolumn{1}{c|}{$\sum_{n=3}^{N-1}=N-3$}
  &\multicolumn{1}{c|}{$c^B_{r,s}$, $(3 \leq n \leq N-1)$}
  & \multicolumn{1}{c|}{$\sum_{n=3}^{N-1}=N-3$}\\
  \cline{1-1}\cline{2-2}\cline{3-3}\cline{4-4}
    \end{tabular*}
    \end{table*}
Then, the total number of equations for describing the system as a
function of maximum chain length $N$ is:
\begin{eqnarray}
\#eqs&=&6+2(N-1)+2(N-2)+2(N-3)+(N^2-7N+12)+2(N-3)=N(N+1),\nonumber\\
\end{eqnarray}
as pointed out in Ref \cite{BH}. From the computational point of
view, the number of equations grows quadratically with the maximum
chain length $N$.

\section{Numerical Results}

We are interested in applying our copolymerization model to fit the
experimental data measured by the Rehovot group, so our primary goal
is to reproduce as closely as possible the details reported
concerning the experiments on chiral amplification of oligopeptides.
For this purpose, the first step is to determine the initial monomer
concentrations to be employed in the simulations. The actual
experiments were carried out for $0.5 mM$ solutions of monomers,
thus we have employed for each case: (a) $R:S=1:1$ which corresponds
to an initial enantiomeric excess $ee_0=0\%$, so
$c^A_{1,0}(0)=0.25mM$ and $c^B_{0,1}(0)=0.25mM$; (b) $R:S=4:6$
corresponding to $ee_0=20\%$, so $c^A_{1,0}(0)=0.2mM$ and
$c^B_{0,1}(0)=0.3mM$; (c) $R:S=3:7$ which corresponds to
$ee_0=40\%$, so $c^A_{1,0}(0)=0.15mM$ and $c^B_{0,1}(0)=0.35mM$. The
remainder of the initial concentrations (the dimers and on up) are
taken to be zero. Next, we systematically search for the reaction
rates leading to the best fit to the given data.

Different chemical model systems were used in the experiments:
namely $\gamma$-stearyl-glutamic thioethyl ester $(C_{18}-TE-Glu)$,
$N^{\epsilon}$-stearoyl-lysine thioethyl ester $(C_{18}-TE-Lys)$,
$\gamma$-stearyl-glutamic acid N-carboxyanhydride $(C_{18}-Glu-NCA)$
and $\gamma$-stearyl-glutamic thioacid $(C_{18}-thio-Glu)$, varying
both their initial compositions and for various choices of catalyst.
The composition of the oligopeptides formed was analyzed by
matrix-assisted laser desorption/ionization time-of-flight mass
spectroscopy (MALDI-TOF) with enantio-labeled samples. The
experimental relative abundances of the oligopeptides was inferred
from the ion intensity. It are these relative abundances that we aim
to interpret vis-a-vis our copolymerization model.

Since only the experiments with racemic mixtures of the starting
compounds required a catalyst, it is reasonable to expect that the
racemic and the chiral enriched cases will follow different dynamics
for a given model system. That is, the presence or absence of a
specific catalyst affects the rate constants, for a given chemical
system. Firstly, we will find the reaction rates for the racemic
case, and afterwards, those for the enriched chiral case, allowing
us to compare both. The a-priori nine free parameters we must set to
run the numerical integrations are comprised by the four direct and
the four inverse rate constants $k_{aa}$, $k_{bb}$, $k_{ab}$,
$k_{ba}$, and $k_{aa}^*$, $k_{bb}^*$, $k_{ab}^*$, $k_{ba}^*$, plus
the maximum polymer chain length, $N$. We set all the inverse
reaction rates to a unique value,
$k_{aa}^*=k_{bb}^*=k_{ab}^*=k_{ba}^* = 10^{-10}(s^{-1})$, implying
an almost irreversible scheme,  and we determine the remainder of
the parameters from fitting the copolymerization model to the
relative abundance data. This required numerical integration of the
set of differential equations Eqs. ({\ref{dA}-\ref{dhetero3maxB})
which we performed using the Mathematica program package. For each
independent run we verified the compliance of the numerical results
with the constraint in Eq.(\ref{mass}), an imperative for any closed
system.

Results from fitting the model to the data indicate that the maximum
chain length $N$ does not play a significant role, the Pearson
product-moment correlation coefficient, $r$, remains the same for
$N=12,14,16,18,20$, so we will set $N=12$ for all compounds and
cases treated below. Since the number of independent equations
scales as $N^2$, this represents an important reduction on computer
time and the memory used. We note that one is free to scale out the
dependence of one pair of reaction constants from the rate equations
by a suitable redefinition of the time variable. Thus, without loss
of generality, we set the cross inhibition rates equal to unity
$k_{ab}=k_{ba}=1 (s^{-1}mol^{-1})$ and then search for the reaction
rates $k_{aa}=k_{bb}$ leading to the best fits.

\subsection{Racemic mixtures}

In one set of experiments, the authors reported MALDI-TOF analysis
of the oligopeptides formed at the air-water interface from racemic
mixtures $R:S = 1:1$ of the monomers for the various model systems
and catalysts. We first fit the copolymerization model to this data.

The best correlation data for the racemic $C_{18}$-TE-Glu system,
with the $I_2/KI$ catalyst are found for
$k_{aa}=k_{bb}=1.7(s^{-1}mol^{-1})$. In this case, the best fit
obtains for the time scale $t=10^{11} (s)$. Exactly by the same
process, the best correlation data for the racemic $C_{18}-TE-Lys$
are found for $k_{aa}=k_{bb}=2.3 (s^{-1}mol^{-1})$ and for
$k_{aa}=k_{bb}=1.3 (s^{-1}mol^{-1})$ when adding $I_2/KI$ and
$AgNO_3$ as catalyst, respectively. For the simulations here, we
took the times $t=10^{10} (s)$ and $t=10^{11} (s)$ in the racemic
cases with $I_2/KI$ and $AgNO_3$ respectively. Finally, we fit our
copolymerization model to the $C_{18}-thio-Glu$ experimental
relative abundances. The authors of the experiments affirmed that
this compound undergoes a truly random polymerization, so fits from
our model are expected to be slightly less satisfactory than those
for the binomial distribution function. Setting the inverse reaction
rates and the cross inhibition as indicated above, then the best
correlation coefficients are found for $k_{aa}=k_{bb}=0.4
(s^{-1}mol^{-1})$. The instant or time-scale leading to these
numerical values is $t=10^{10} (s)$.

The corresponding (experimental and numerical) relative abundances
for the four compounds cited above corresponding to these values are
shown in Fig. \ref{allracemic}. The histograms show the relative
abundance of each experimentally obtained oligopeptide compared to
the best fit from our copolymerization model.  We emphasize that we
fit the model to the complete family of stereoisomer subgroups
(global fit). The resulting data correlations are shown in Fig.
\ref{fitallracemic} and Table \ref{racemic}, the latter gives a
detailed comparison of the best fits between individual subfamilies
and the overall global fit.

\begin{figure}[h!]
\centering
  \includegraphics[width=0.5\textwidth]{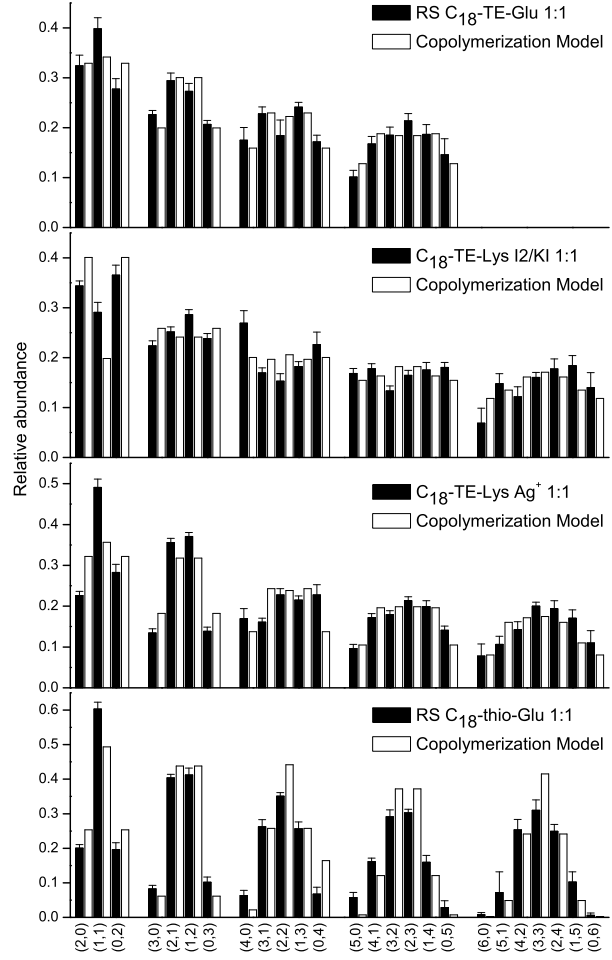}
  \caption{Relative abundance versus number of repeat units $(r,s)$ of the oligopeptides obtained from fitting
  the model (white) to the experimental data (black) from racemic mixtures $R:S=1:1$ of monomers. The four chemical models
  are indicated by the insets.}
  \label{allracemic}
\end{figure}

\begin{figure}[h!]
\centering
  \includegraphics[width=0.5\textwidth]{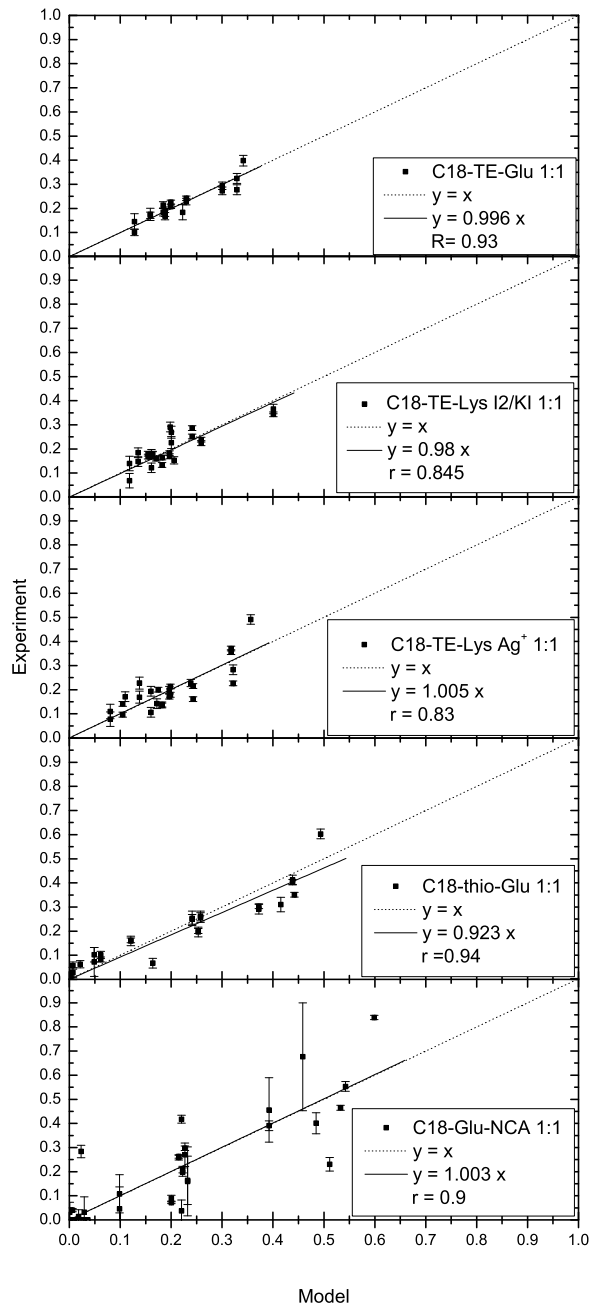}
  \caption{Data correlations r from fitting the model to the data in the case
  of racemic mixtures of all the compounds employed. The chemical systems are indicated by the insets.
  The solid line represents the linear
  correlation between experimental data and numerical calculations.}
  \label{fitallracemic}
\end{figure}
\begin{table}[h!]
\centering \small
  \caption{\ Comparative fits between the copolymerization model and the binomial distribution to the experimental relative abundances: racemic
  mixtures $R:S=1:1$ of monomers of the four model systems as indicated in the leftmost column. Only in the case of $C_{18}-thio-Glu$ does the binomial distribution give a better global fit than
  the copolymerization model: this latter system provides an experimental reference system for random polymerization \cite{Lahav2003b}. }
  \label{racemic}
    \renewcommand{\arraystretch}{1.5}
    \renewcommand\tabcolsep{1pt}
\begin{tabular*}{0.5\textwidth}{ c | c | c | c | c| c| c| c|} \cline{2-2}\cline{3-3}\cline{4-4}\cline{5-5}\cline{6-6}\cline{7-7}\cline{8-8}
  & \multicolumn{6}{c|}{\textbf{Copolymerization model}}
  & \multicolumn{1}{c|}{\textcolor[rgb]{0.50,0.50,0.50}{\textbf{Bin.}}} \\
  \cline{2-2}\cline{3-3}\cline{4-4}\cline{5-5}\cline{6-6}\cline{7-7}\cline{8-8}
  \textbf{$r$} & \multicolumn{5}{c|}{\textbf{Fits for each subgroup n}}
  & \multicolumn{1}{c|}{\textbf{Global}}
  & \multicolumn{1}{c|}{\textcolor[rgb]{0.50,0.50,0.50}{\textbf{Global}}}\\
  \cline{2-2}\cline{2-2}\cline{3-3}\cline{4-4}\cline{5-5}\cline{6-6}
  \multicolumn{1}{c|}{}
  & \textbf{di}
  & \textbf{tri}
  & \textbf{tetra}
  & \textbf{penta}
  & \textbf{hexa}
  & \multicolumn{1}{c|}{\textbf{fit}}
  & \multicolumn{1}{c|}{\textcolor[rgb]{0.50,0.50,0.50}{\textbf{fit}}}\\
  \cline{1-1}\cline{2-2}\cline{3-3}\cline{4-4}\cline{5-5}\cline{6-6}\cline{7-7}\cline{8-8}
  \multicolumn{1}{|c|}{$C_{18}-TE-Glu$}
  & 0.92
  & 0.96
  & 0.80
  & 0.84
  & -
  & 0.93
  & \textcolor[rgb]{0.50,0.50,0.50}{0.75}\\
  \cline{1-1}\cline{2-2}\cline{3-3}\cline{4-4}\cline{5-5}\cline{6-6}\cline{7-7}\cline{8-8}
  \multicolumn{1}{|c|}{$C_{18}-TE-Lys (I_2/KI)$}
  & 0.96
  & -0.82
  & -0.11
  & -0.73
  & 0.45
  & 0.85
  & \textcolor[rgb]{0.50,0.50,0.50}{0.32}\\
  \cline{1-1}\cline{2-2}\cline{3-3}\cline{4-4}\cline{5-5}\cline{6-6}\cline{7-7}\cline{8-8}
  \multicolumn{1}{|c|}{$C_{18}-TE-Lys (Ag)$}
  & 0.98
  & 1
  & 0.03
  & 0.88
  & 0.76
  & 0.84
  & \textcolor[rgb]{0.50,0.50,0.50}{0.8}\\
  \cline{1-1}\cline{2-2}\cline{3-3}\cline{4-4}\cline{5-5}\cline{6-6}\cline{7-7}\cline{8-8}
  \multicolumn{1}{|c|}{$C_{18}-thio-Glu$}
  & 1
  & 1
  & 1
  & 0.98
  & 0.97
  & 0.95
  & \textcolor[rgb]{0.50,0.50,0.50}{0.98}\\
  \cline{1-1}\cline{2-2}\cline{3-3}\cline{4-4}\cline{5-5}\cline{6-6}\cline{7-7}\cline{8-8}
  \end{tabular*}
    \end{table}
In the case of the $C_{18}-Glu-NCA$ with catalyst $Ni(CH_3CO_2)_2$,
the best fit is obtained for $k_{aa}=k_{bb}=0.2 (s^{-1}mol^{-1})$.
Results for the corresponding relative abundances are shown in Fig.
\ref{racGluNCA} and the correlation from fitting is displayed in the
bottom frame of Fig.\ref{fitallracemic} and Table
\ref{racemicGluNCA}. Not all subfamily data sets are reported in the
experimental paper \cite{Lahav2002b}; here we use the fitted model
to the partial data set to predict or fill in this missing subfamily
data. Numerical results for the racemic case
have been found for $t=10^{10} (s)$.\\
\begin{figure*}[h!]
  \includegraphics[height=6.7cm]{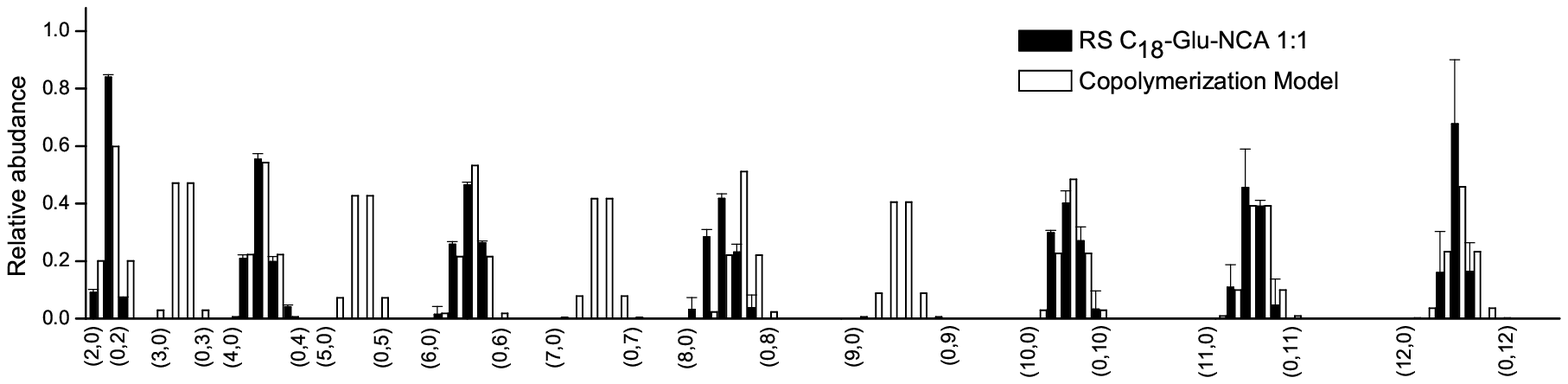}
  \caption{The $C_{18}-Glu-NCA$ system with catalyst $Ni(CH_3CO_2)_2$: relative abundance of the oligopeptides obtained from fitting
  the model (white) to the experimental data (black) from racemic mixtures of monomers.
  Compare to Fig. 4A of reference\cite{Lahav2003b}}
  \label{racGluNCA}
\end{figure*}

\begin{center}
\begin{table}[h!] \small
  \caption{\ Comparative fits between the copolymerization model and binomial distribution to the experimental relative abundances for
  the racemic compositions (R:S=1:1) of the $C_{18}-Glu-NCA$ system.}
  \label{racemicGluNCA}
    \renewcommand{\arraystretch}{1.5}
    \renewcommand\tabcolsep{3pt}
\begin{tabular*}{0.9\textwidth}{ c | c | c | c | c| c| c| c|c| c| c| c| c|c|} \cline{2-2}\cline{3-3}\cline{4-4}\cline{5-5}\cline{6-6}\cline{7-7}\cline{8-8}\cline{9-9}\cline{10-10}\cline{11-11}\cline{12-12}\cline{13-13}\cline{14-14}
  & \multicolumn{12}{c|}{\textbf{Copolymerization model}}
  & \multicolumn{1}{c|}{\textcolor[rgb]{0.50,0.50,0.50}{\textbf{Bin.}}} \\
  \cline{2-2}\cline{3-3}\cline{4-4}\cline{5-5}\cline{6-6}\cline{7-7}\cline{8-8}\cline{9-9}\cline{10-10}\cline{11-11}\cline{12-12}\cline{13-13}\cline{14-14}
  \textbf{$r$} & \multicolumn{11}{c|}{\textbf{Fits for each subgroup n}}
  & \multicolumn{1}{c|}{\textbf{Global}}
  & \multicolumn{1}{c|}{\textcolor[rgb]{0.50,0.50,0.50}{\textbf{Global}}}\\
  \cline{2-2}\cline{2-2}\cline{3-3}\cline{4-4}\cline{5-5}\cline{6-6}\cline{7-7}\cline{8-8}\cline{9-9}\cline{10-10}\cline{11-11}\cline{12-12}
  \multicolumn{1}{c|}{}
  & \textbf{di}
  & \textbf{tri}
  & \textbf{tetra}
  & \textbf{penta}
  & \textbf{hexa}
  & \textbf{hepta}
  & \textbf{octa}
  & \textbf{nona}
  & \textbf{deca}
  & \textbf{endeca}
  & \textbf{dodeca}
  & \multicolumn{1}{c|}{\textbf{fit}}
  & \multicolumn{1}{c|}{\textcolor[rgb]{0.50,0.50,0.50}{\textbf{fit}}}\\
  \cline{1-1}\cline{2-2}\cline{3-3}\cline{4-4}\cline{5-5}\cline{6-6}\cline{7-7}\cline{8-8}\cline{9-9}\cline{10-10}\cline{11-11}\cline{12-12}\cline{13-13}\cline{14-14}
  \multicolumn{1}{|c|}{$C_{18}-Glu-NCA$}
  & 1
  & -
  & 1
  & -
  & 0.98
  & -
  & 0.98
  & -
  & 0.97
  & 1
  & 0.95
  & 0.96
  & \textcolor[rgb]{0.50,0.50,0.50}{0.75}\\
 \cline{1-1} \cline{2-2}\cline{3-3}\cline{4-4}\cline{5-5}\cline{6-6}\cline{7-7}\cline{8-8}\cline{9-9}\cline{10-10}\cline{11-11}\cline{12-12}\cline{13-13}\cline{14-14}
  \end{tabular*}
    \end{table}
    \end{center}

\subsection{Chirally enriched mixtures}

In a second set of experiments, the authors reported MALDI-TOF
analysis of the oligopeptides formed at the air-water interface from
non-racemic mixtures of the monomers for the same model systems. No
catalysts were employed there. We next consider fits of our model to
these data sets.

The best correlations factors for both chirally enriched mixture
cases (20\% and 40\% excesses) in the case of the $C_{18}-TE-Glu$
system are found for the same rates, that is for $k_{aa}=k_{bb}=2
(s^{-1}mol^{-1})$. The results for these values are shown in Table
\ref{rchTEGlu}. In Fig. \ref{chTEGlu} we display the relative
abundances of the homochiral oligopeptides and in Table
\ref{chTEGluee} both the calculated and experimental enantiomeric
excesses for the 4:6 and 3:7 (R:S) mixtures. In Fig.
\ref{chfitTEGlu} we show the data correlation. Numerical results for
the non-racemic case have been found for the time scale $t=10^{11}
(s)$.

\begin{table}[h!] \centering \small
  \caption{\ Comparative fits between the copolymerization model and the binomial distribution to
  the experimental relative abundances measured for
  non-racemic mixtures of $C_{18}-TE-Glu$.}
  \label{rchTEGlu}
    \renewcommand{\arraystretch}{1.5}
    \renewcommand\tabcolsep{3pt}
\begin{tabular*}{0.5\textwidth}{ c | c | c | c | c| c| c| c|} \cline{2-2}\cline{3-3}\cline{4-4}\cline{5-5}\cline{6-6}\cline{7-7}\cline{8-8}
  & \multicolumn{6}{c|}{\textbf{Copolymerization model}}
  & \multicolumn{1}{c|}{\textcolor[rgb]{0.50,0.50,0.50}{\textbf{Binomial}}} \\
  \cline{2-2}\cline{3-3}\cline{4-4}\cline{5-5}\cline{6-6}\cline{7-7}\cline{8-8}
  \textbf{$r$} & \multicolumn{5}{c|}{\textbf{Fits for each subgroup n}}
  & \multicolumn{1}{c|}{\textbf{Global}}
  & \multicolumn{1}{c|}{\textcolor[rgb]{0.50,0.50,0.50}{\textbf{Global}}}\\
  \cline{2-2}\cline{2-2}\cline{3-3}\cline{4-4}\cline{5-5}\cline{6-6}
  \multicolumn{1}{c|}{}
  & \textbf{di}
  & \textbf{tri}
  & \textbf{tetra}
  & \textbf{penta}
  & \textbf{hexa}
  & \multicolumn{1}{c|}{\textbf{fit}}
  & \multicolumn{1}{c|}{\textcolor[rgb]{0.50,0.50,0.50}{\textbf{fit}}}\\
  \cline{1-1}\cline{2-2}\cline{3-3}\cline{4-4}\cline{5-5}\cline{6-6}\cline{7-7}\cline{8-8}
  \multicolumn{1}{|c|}{$\textbf{(R:S)\, 4:6}$}
  & 0.86
  & 0.89
  & 0.93
  & 0.99
  & -
  & 0.94
  & \textcolor[rgb]{0.50,0.50,0.50}{0.75}\\
  \cline{1-1}\cline{2-2}\cline{3-3}\cline{4-4}\cline{5-5}\cline{6-6}\cline{7-7}\cline{8-8}
  \multicolumn{1}{|c|}{$\textbf{(R:S)\, 3:7}$}
  & 0.95
  & 0.94
  & 0.96
  & 0.99
  & 0.99
  & 0.95
  & \textcolor[rgb]{0.50,0.50,0.50}{0.85}\\
  \cline{1-1}\cline{2-2}\cline{3-3}\cline{4-4}\cline{5-5}\cline{6-6}\cline{7-7}\cline{8-8}
  \end{tabular*}
    \end{table}

\begin{table}[h!]
\centering \small
  \caption{\ Enantiomeric excesses $ee$: numerical results from the copolymerization model (experimental data)
  for the relative abundances of the homochiral oligopeptides for the $C_{18}-TE-Glu$ system.}
  \label{chTEGluee}
    \renewcommand{\arraystretch}{1.5}
    \renewcommand\tabcolsep{4.6pt}
\begin{tabular*}{0.5\textwidth}{ c | c | c | c | c| c| }
\cline{2-2}\cline{3-3}\cline{4-4}\cline{5-5}\cline{6-6}
    \cline{2-2}\cline{2-2}\cline{3-3}\cline{4-4}\cline{5-5}\cline{6-6}
  \multicolumn{1}{c|}{\textbf{$ee(\%)$}}
  & \textbf{di}
  & \textbf{tri}
  & \textbf{tetra}
  & \textbf{penta}
  & \textbf{hexa}\\
  \cline{1-1}\cline{2-2}\cline{3-3}\cline{4-4}\cline{5-5}\cline{6-6}
  \multicolumn{1}{|c|}{$\textbf{(R:S)\, 4:6}$}
  & 18 (26)
  & 24 (39)
  & 30 (46)
  & 35 (59)
  & -\\
  \cline{1-1}\cline{2-2}\cline{3-3}\cline{4-4}\cline{5-5}\cline{6-6}
  \multicolumn{1}{|c|}{$\textbf{(R:S)\, 3:7}$}
  & 37 (48)
  & 48 (71)
  & 57 (82)
  & 66 (92)
  & 73 ($>$99.8)\\
  \cline{1-1}\cline{2-2}\cline{3-3}\cline{4-4}\cline{5-5}\cline{6-6}
  \end{tabular*}
    \end{table}

\begin{figure}[h!]
\centering
  \includegraphics[width=0.5\textwidth]{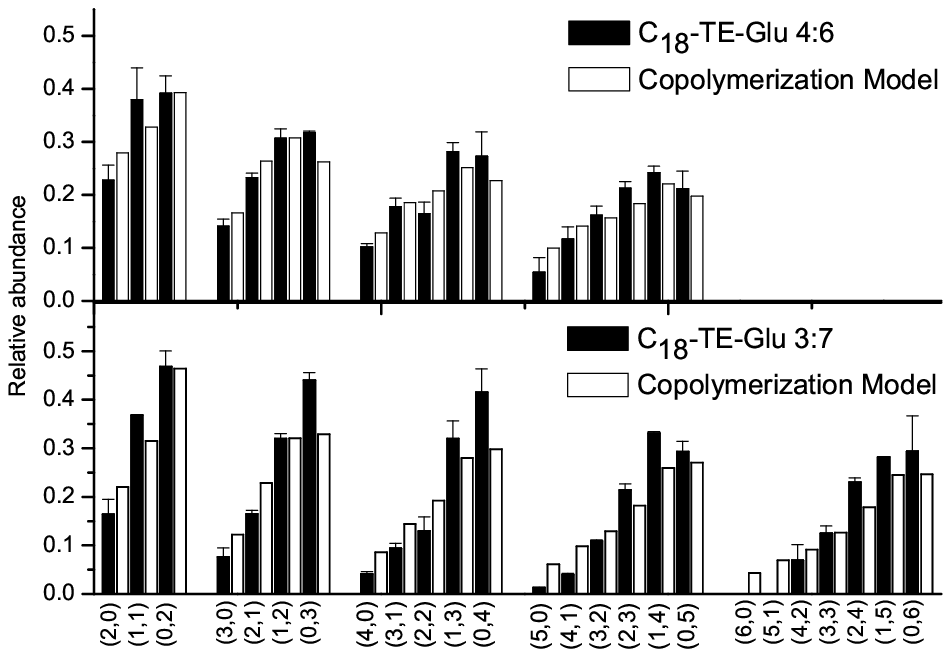}
  \caption{Relative abundance versus number of repeat units $(r,s)$ of the oligopeptides obtained from fitting
  the model (white) to the experimental data (black) from non-racemic mixtures of monomers for the $C_{18}-TE-Glu$ system.}
  \label{chTEGlu}
\end{figure}

\begin{figure}[h!]
\centering
  \includegraphics[width=0.5\textwidth]{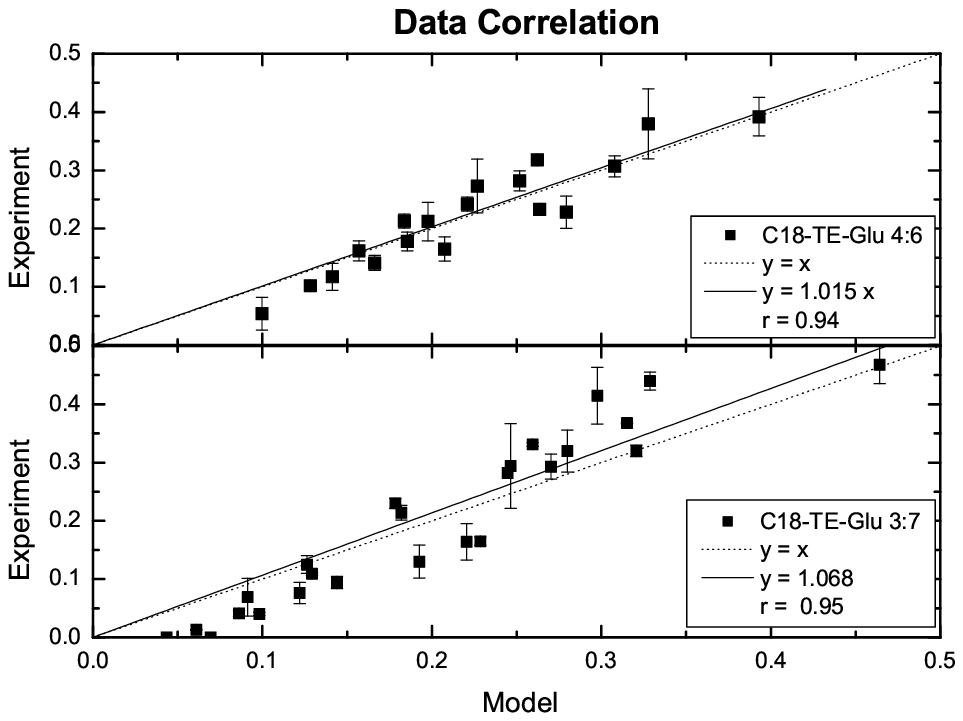}
  \caption{Results from fitting the model to the experimental data:
  non-racemic mixtures of $C_{18}-TE-Glu$.
  The solid line represents the linear correlation between experimental and numerical data obtained from fitting.
  The dotted line has slope equal to unity.}
  \label{chfitTEGlu}
\end{figure}
For the chiral mixtures of $C_{18}-TE-Lys$ we found the best fits
for the dynamics corresponding to $k_{aa}=k_{bb}=2.5
(s^{-1}mol^{-1})$. The results for these values are shown in Table
\ref{rchTELys}. The relative abundances results for these values are
shown in Fig. \ref{chTELys} and the enantiomeric excesses obtained
for 4:6 and 3:7 (R:S) mixtures are presented in Table
\ref{chTELysee}. In Fig. \ref{chfitTELys} the data correlation is
shown. For the simulations here, we took the instants $t=10^{10}
(s)$ and $t=10^{11} (s)$ in the racemic cases with $I_2/KI$ and
$AgNO_3$ respectively, and $t=10^{10} (s)$ for the chirally enriched
mixtures.
\begin{table}[h!]
\centering \small
  \caption{\ Results for the copolymerization model and experimental data correlations for
  non-racemic mixtures of $C_{18}-TE-Lys$.}
  \label{rchTELys}
    \renewcommand{\arraystretch}{1.5}
    \renewcommand\tabcolsep{1.5pt}
\begin{tabular*}{0.5\textwidth}{ c | c | c | c | c| c| c| c|c|} \cline{2-2}\cline{3-3}\cline{4-4}\cline{5-5}\cline{6-6}\cline{7-7}\cline{8-8}\cline{9-9}
  & \multicolumn{7}{c|}{\textbf{Copolymerization model}}
  & \multicolumn{1}{c|}{\textcolor[rgb]{0.50,0.50,0.50}{\textbf{Binomial}}} \\
  \cline{2-2}\cline{3-3}\cline{4-4}\cline{5-5}\cline{6-6}\cline{7-7}\cline{8-8}\cline{9-9}
  \textbf{$r$} & \multicolumn{6}{c|}{\textbf{Fits for each subgroup n}}
  & \multicolumn{1}{c|}{\textbf{Global}}
  & \multicolumn{1}{c|}{\textcolor[rgb]{0.50,0.50,0.50}{\textbf{Global}}}\\
  \cline{2-2}\cline{2-2}\cline{3-3}\cline{4-4}\cline{5-5}\cline{6-6}\cline{7-7}
  \multicolumn{1}{c|}{}
  & \textbf{di}
  & \textbf{tri}
  & \textbf{tetra}
  & \textbf{penta}
  & \textbf{hexa}
  & \textbf{hepta}
  & \multicolumn{1}{c|}{\textbf{fit}}
  & \multicolumn{1}{c|}{\textcolor[rgb]{0.50,0.50,0.50}{\textbf{fit}}}\\
  \cline{1-1}\cline{2-2}\cline{3-3}\cline{4-4}\cline{5-5}\cline{6-6}\cline{7-7}\cline{8-8}\cline{9-9}
  \multicolumn{1}{|c|}{$\textbf{(R:S)\, 4:6}$}
  & 0.78
  & 1
  & 0.87
  & 0.90
  & 0.84
  & 0.97
  & 0.89
  & \textcolor[rgb]{0.50,0.50,0.50}{0.46}\\
  \cline{1-1}\cline{2-2}\cline{3-3}\cline{4-4}\cline{5-5}\cline{6-6}\cline{7-7}\cline{8-8}\cline{9-9}
  \multicolumn{1}{|c|}{$\textbf{(R:S)\, 3:7}$}
  & 0.93
  & 1
  & 0.95
  & 0.97
  & 0.99
  & -
  & 0.94
  & \textcolor[rgb]{0.50,0.50,0.50}{0.65}\\
  \cline{1-1}\cline{2-2}\cline{3-3}\cline{4-4}\cline{5-5}\cline{6-6}\cline{7-7}\cline{8-8}\cline{9-9}
  \end{tabular*}
    \end{table}

\begin{table}[h!]
\centering \small
  \caption{\ Enantiomeric excesses: numerical results from the copolymerization model (experimental data)
  for the relative abundances of the homochiral oligopeptides for $C_{18}-TE-Lys$.}
  \label{chTELysee}
    \renewcommand{\arraystretch}{1.5}
    \renewcommand\tabcolsep{2.2pt}
\begin{tabular*}{0.5\textwidth}{ c | c | c | c | c| c|c| }
\cline{2-2}\cline{3-3}\cline{4-4}\cline{5-5}\cline{6-6}
    \cline{2-2}\cline{2-2}\cline{3-3}\cline{4-4}\cline{5-5}\cline{6-6}\cline{7-7}
  \multicolumn{1}{c|}{\textbf{$ee(\%)$}}
  & \textbf{di}
  & \textbf{tri}
  & \textbf{tetra}
  & \textbf{penta}
  & \textbf{hexa}
  & \textbf{hepta}\\
  \cline{1-1}\cline{2-2}\cline{3-3}\cline{4-4}\cline{5-5}\cline{6-6}\cline{7-7}
  \multicolumn{1}{|c|}{$\textbf{(R:S)\, 4:6}$}
  & 23 (34)
  & 30 (34)
  & 36 (41)
  & 42 (60)
  & 49 (62)
  & 54 ($>$99.8)\\
  \cline{1-1}\cline{2-2}\cline{3-3}\cline{4-4}\cline{5-5}\cline{6-6}\cline{7-7}
  \multicolumn{1}{|c|}{$\textbf{(R:S)\, 3:7}$}
  & 45 (46)
  & 57 (63)
  & 66 (73)
  & 75 (85)
  & 81 (86)
  & - \\
  \cline{1-1}\cline{2-2}\cline{3-3}\cline{4-4}\cline{5-5}\cline{6-6}\cline{7-7}
  \end{tabular*}
    \end{table}

\begin{figure}[h!]
\centering
  \includegraphics[width=0.5\textwidth]{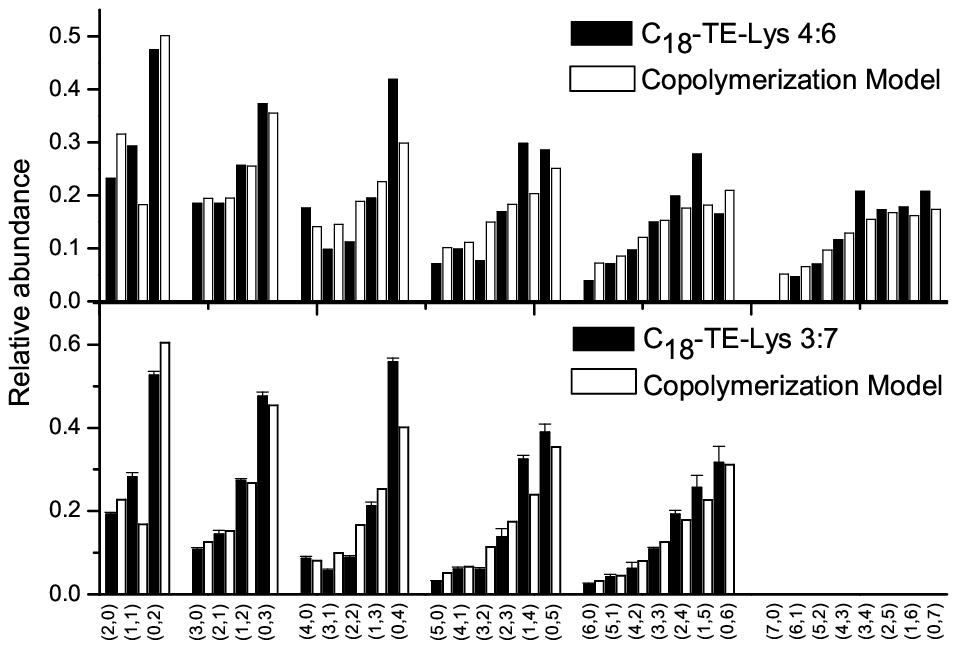}
  \caption{Relative abundance versus number of repeat units $(r,s)$ of the oligopeptides obtained from fitting
  the model (white) to the experimental data (black) from non-racemic mixtures of monomers of $C_{18}-TE-Lys$. }
  \label{chTELys}
\end{figure}

\begin{figure}[h!]
\centering
  \includegraphics[width=0.5\textwidth]{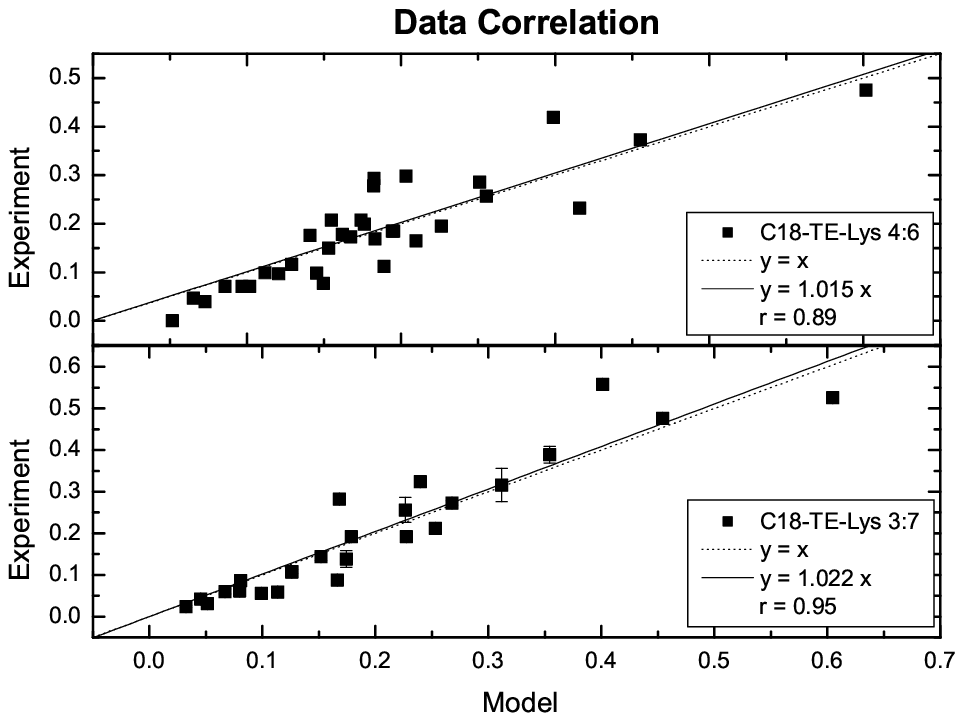}
  \caption{Results from fitting the model to the experimental data.
  Chiral mixtures of $C_{18}-TE-Lys$.
  The solid line represents the linear correlation between experimental and numerical data obtained from fitting.
  The dotted line has slope equal to unity.}
  \label{chfitTELys}
\end{figure}
In the case of nonracemic $C_{18}-thio-Glu$, the best correlation
coefficients are found for the same values of the reaction rates
that we found in the racemic case, namely for  $k_{aa}=k_{bb}=0.4
(s^{-1}mol^{-1})$. Results for the chiral cases are shown in Table
\ref{rchthioGlu}. As to be expected and as shown there, the
correlation factors for the global fit to the binomial distribution
function are slighter better than those for any simulation we could
perform with the copolymerization model, so we reconfirm what was
claimed by the authors of the experimental work: namely that the
$C_{18}-thio-Glu$ system polymerizes randomly. In Figure
\ref{chthioGlu} the relative abundances of the oligopeptides are
shown.  The data correlation is shown in Fig.\ref{chfitthioGlu}.

\begin{table}[h!]
\centering \small
  \caption{\ Results for the copolymerization model and experimental data correlations for non-racemic mixtures of $C_{18}-thio-Glu$.}
  \label{rchthioGlu}
    \renewcommand{\arraystretch}{1.5}
    \renewcommand\tabcolsep{1.5pt}
\begin{tabular*}{0.5\textwidth}{ c | c | c | c | c| c| c| c|c|} \cline{2-2}\cline{3-3}\cline{4-4}\cline{5-5}\cline{6-6}\cline{7-7}\cline{8-8}\cline{9-9}
  & \multicolumn{7}{c|}{\textbf{Copolymerization model}}
  & \multicolumn{1}{c|}{\textcolor[rgb]{0.50,0.50,0.50}{\textbf{Binomial}}} \\
  \cline{2-2}\cline{3-3}\cline{4-4}\cline{5-5}\cline{6-6}\cline{7-7}\cline{8-8}\cline{9-9}
  $\textbf{r}$ & \multicolumn{6}{c|}{\textbf{Fits for each subgroup n}}
  & \multicolumn{1}{c|}{\textbf{Global}}
  & \multicolumn{1}{c|}{\textcolor[rgb]{0.50,0.50,0.50}{\textbf{Global}}}\\
  \cline{2-2}\cline{2-2}\cline{3-3}\cline{4-4}\cline{5-5}\cline{6-6}\cline{7-7}
  \multicolumn{1}{c|}{}
  & \textbf{di}
  & \textbf{tri}
  & \textbf{tetra}
  & \textbf{penta}
  & \textbf{hexa}
  & \textbf{hepta}
  & \multicolumn{1}{c|}{\textbf{fit}}
  & \multicolumn{1}{c|}{\textcolor[rgb]{0.50,0.50,0.50}{\textbf{fit}}}\\
  \cline{1-1}\cline{2-2}\cline{3-3}\cline{4-4}\cline{5-5}\cline{6-6}\cline{7-7}\cline{8-8}\cline{9-9}
  \multicolumn{1}{|c|}{$\textbf{(R:S)\, 4:6}$}
  & 0.93
  & 0.98
  & 0.93
  & 0.92
  & 0.92
  & 0.91
  & 0.91
  & \textcolor[rgb]{0.50,0.50,0.50}{0.93}\\
  \cline{1-1}\cline{2-2}\cline{3-3}\cline{4-4}\cline{5-5}\cline{6-6}\cline{7-7}\cline{8-8}\cline{9-9}
  \multicolumn{1}{|c|}{$\textbf{(R:S)\, 3:7}$}
  & 0.89
  & 1
  & 0.99
  & 0.99
  & 0.98
  & -
  & 0.96
  & \textcolor[rgb]{0.50,0.50,0.50}{0.97}\\
  \cline{1-1}\cline{2-2}\cline{3-3}\cline{4-4}\cline{5-5}\cline{6-6}\cline{7-7}\cline{8-8}\cline{9-9}
  \end{tabular*}
    \end{table}

\begin{figure}[h!]
\centering
  \includegraphics[width=0.5\textwidth]{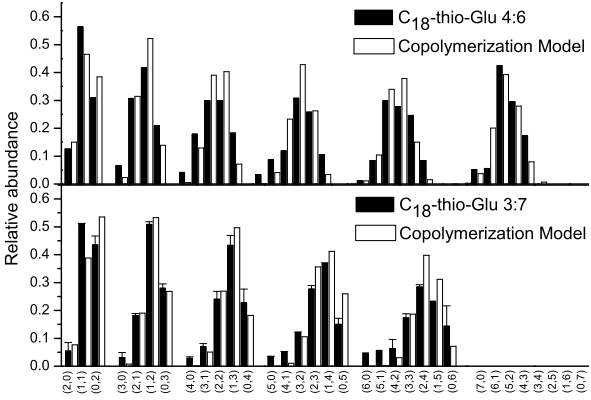}
  \caption{Relative abundances versus number of repeat units $(r,s)$ of the oligopeptides obtained from fitting
  the model (white) to the experimental data (black) for the non-racemic mixtures of $C_{18}-thio-Glu$.}
  \label{chthioGlu}
\end{figure}

\begin{figure}[h!]
\centering
  \includegraphics[width=0.5\textwidth]{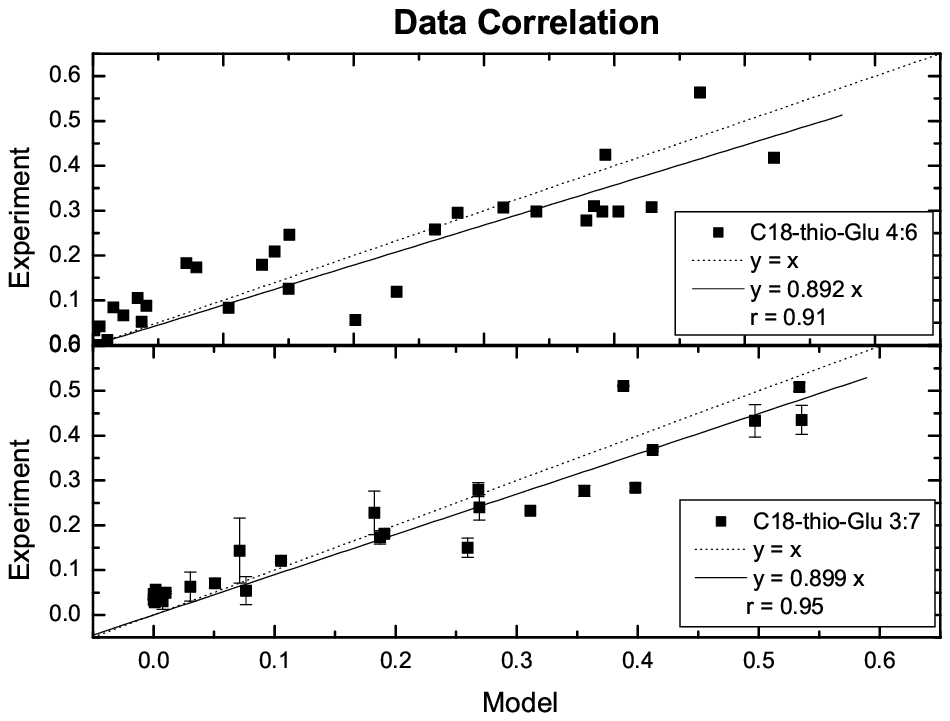}
  \caption{Results from fitting the model to the experimental data for non-racemic mixtures of the $C_{18}-thio-Glu$ system.
  The solid line represents the linear correlation between experimental and numerical data obtained from fitting.
  The dotted line has slope equal to unity.}
  \label{chfitthioGlu}
\end{figure}

The best fits for both chirally enriched mixture cases (20\% and
40\% excesses) in the case of $C_{18}-Glu-NCA$ are found for the
same dynamics, that is $k_{aa}=k_{bb}=3.8 (s^{-1}mol^{-1})$. The
results for these values are shown in Table \ref{rchGluNCA}. In
Fig.\ref{chGluNCA} we compare the best fit against the
experimentally obtained relative abundances of the oligopeptides.
The corresponding data correlation is shown in
Fig.\ref{chfitGluNCA}. Numerical results for the racemic case have
been found for $t=10^{11} (s)$.

\begin{table*}[h!]
\centering \small
  \caption{\ Results for the copolymerization model and experimental data correlations for $C_{18}-Glu-NCA$. The global fit from the
  binomial distribution is shown for comparison.}
  \label{rchGluNCA}
    \renewcommand{\arraystretch}{1.5}
    \renewcommand\tabcolsep{3pt}
\begin{tabular*}{0.8\textwidth}{ c | c | c | c | c| c| c| c|c|c|c|c|}
\cline{2-2}\cline{3-3}\cline{4-4}\cline{5-5}\cline{6-6}\cline{7-7}\cline{8-8}\cline{9-9}\cline{10-10}\cline{11-11}\cline{12-12}
  & \multicolumn{10}{c|}{\textbf{Copolymerization model}}
  & \multicolumn{1}{c|}{\textcolor[rgb]{0.50,0.50,0.50}{\textbf{Binomial}}} \\
  \cline{2-2}\cline{3-3}\cline{4-4}\cline{5-5}\cline{6-6}\cline{7-7}\cline{8-8}\cline{9-9}\cline{10-10}\cline{11-11}\cline{12-12}
  $\textbf{r}$ & \multicolumn{9}{c|}{\textbf{Fits for each subgroup n}}
  & \multicolumn{1}{c|}{\textbf{Global}}
  & \multicolumn{1}{c|}{\textcolor[rgb]{0.50,0.50,0.50}{\textbf{Global}}}\\
  \cline{2-2}\cline{2-2}\cline{3-3}\cline{4-4}\cline{5-5}\cline{6-6}\cline{7-7}\cline{8-8}\cline{9-9}\cline{10-10}
  \multicolumn{1}{c|}{}
  & \textbf{di}
  & \textbf{tri}
  & \textbf{tetra}
  & \textbf{penta}
  & \textbf{hexa}
  & \textbf{hepta}
  & \textbf{octa}
  & \textbf{nona}
  & \textbf{deca}
  & \multicolumn{1}{c|}{\textbf{fit}}
  & \multicolumn{1}{c|}{\textcolor[rgb]{0.50,0.50,0.50}{\textbf{fit}}}\\
  \cline{1-1}\cline{2-2}\cline{3-3}\cline{4-4}\cline{5-5}\cline{6-6}\cline{7-7}\cline{8-8}\cline{9-9}\cline{10-10}\cline{11-11}\cline{12-12}
  \multicolumn{1}{|c|}{$\textbf{(R:S)\, 4:6}$}
  & -0.79
  & 0.9
  & 0.63
  & 0.74
  & 0.95
  & 0.89
  & 0.77
  & 0.86
  & -
  & 0.68
  & \textcolor[rgb]{0.50,0.50,0.50}{0.11}\\
  \cline{1-1}\cline{2-2}\cline{3-3}\cline{4-4}\cline{5-5}\cline{6-6}\cline{7-7}\cline{8-8}\cline{9-9}\cline{10-10}\cline{11-11}\cline{12-12}
  \multicolumn{1}{|c|}{$\textbf{(R:S)\, 3:7}$}
  & -0.33
  & 0.79
  & 0.81
  & 0.76
  & 0.86
  & 0.96
  & 0.75
  & 0.83
  & 0.89
  & 0.75
  & \textcolor[rgb]{0.50,0.50,0.50}{0.38}\\
  \cline{1-1}\cline{2-2}\cline{3-3}\cline{4-4}\cline{5-5}\cline{6-6}\cline{7-7}\cline{8-8}\cline{9-9}\cline{10-10}\cline{11-11}\cline{12-12}
  \end{tabular*}
    \end{table*}

\begin{figure*}[h!]
\centering
  \includegraphics[width=\textwidth]{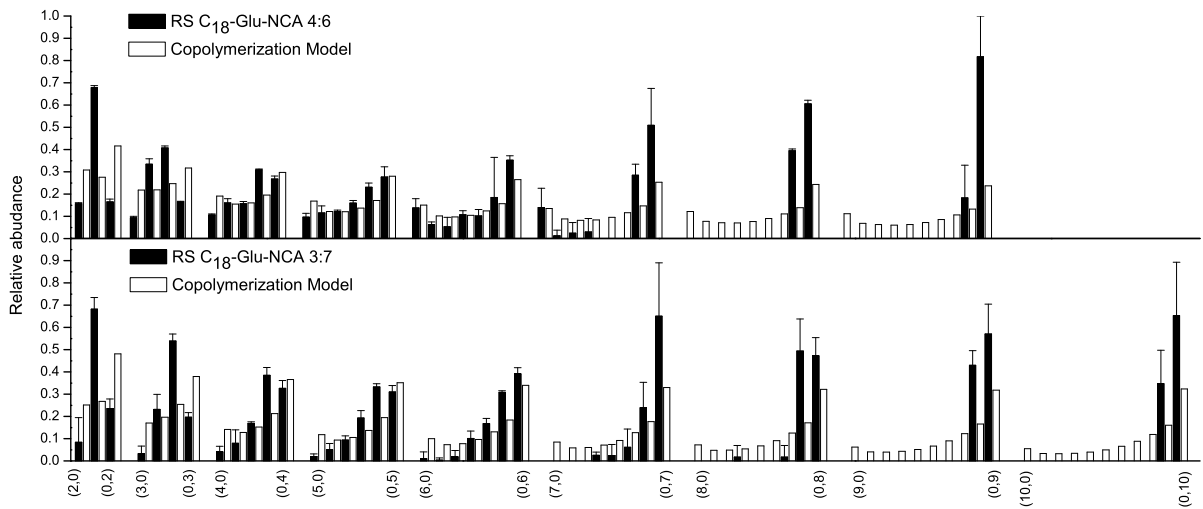}
   \caption{Relative abundances for the non racemic mixtures of $C_{18}-thio-Glu$. The experimental data set (black) is incomplete, we have used our model
   to fill in the missing portions of the histogram (white)}
\label{chGluNCA}
\end{figure*}

\begin{figure}[h!]
\centering
  \includegraphics[width=0.5\textwidth]{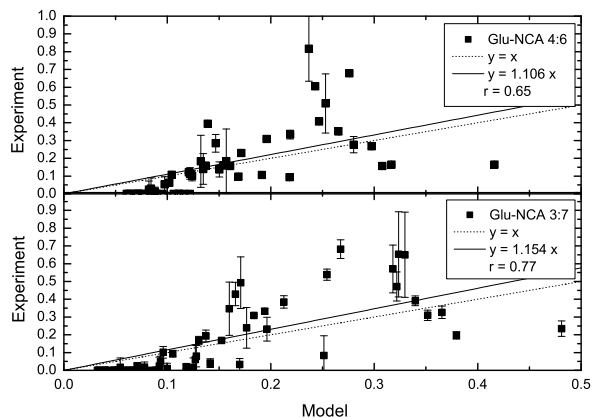}
  \caption{Results from fitting the model to the experimental data for the $C_{18}-Glu-NCA$ system.
  The solid line represents the linear correlation between experimental and numerical data obtained from fitting.
  The dotted line has slope equal to unity.}
  \label{chfitGluNCA}
\end{figure}

\section{Conclusions}

The overall scheme for the chiral amplification process leading to
the experimental data investigated here involves a self-assembly
step followed by a lattice-controlled polymerization
\cite{Lahav2002b,Lahav2003b}. It is this subsequent polymerization
which is the prime focus of this paper. The authors of the
experimental work stress that it is not at all straightforward to
actually establish the \textit{correlation} between the packing
arrangement of the crystallites and the composition of the
diastereoisomeric products that result therefrom. Therefore, our
task here was to fit the outcome of these latter steps assuming an
\textit{effective} copolymerization scheme. The term ``effective"
simply means that the putative complicated correlations and
interplay between the 2D crystallite phases at the air-water
interface and the polymerization reaction pathways that depend on
the microscopic packing arrangements within the crystals are treated
here with a simple model. In this regard, our model can be regarded
as a ``course-grained" description of the overall process in that
the microscopic details (the structures of the crystalline phases)
are not resolved, but that the end-result or net effect of the
pathways afforded by the crystallites can be summarized by the
polymerization scheme as depicted graphically in Fig. \ref{scheme}.

The model as introduced is defined for fully reversible reactions
and this implies that some of the reaction rates must obey a
corresponding constraint as dictated by microreversibility. Thus the
model is appropriate for closed systems under thermodynamic control.
For the numerical fits themselves, we found that all the reverse
reaction rates could be set to rather tiny values, and this in
consonance with experimentally observed irreversible condensation.
Thus for the present purposes, the copolymerization model is
practically irreversible. The values for the forward rates  of
adding the same chirality monomer to the end of the growing chain
are found to be greater than those for addition of a wrong chirality
monomer: that is, $k_{aa} = k_{bb}
> k_{ab}=k_{ba} = 1$ (except of course for the model system $C_{18}-thio-Glu$ serving as
reference for random polymerization).

Other closed systems that lead to copolymers could be in principle
be fit with our model. If for example $k_{aa}$ and $k_{bb}$ had
different magnitudes, this would imply that a underlying chiral bias
is operative either in the polymerization or in the prior formation
of the two crystallites that control the polymerization. This bias
could affect the packing arrangements of the crystal monomers and
the  reaction pathways taken within each crystallite phase. Since
however our model is effective, as explained earlier, we would not
be able to say  whether the chiral bias is in the polymerization or
in the structure of the crystallites that control the
polymerization. Nevertheless, this bias in $k_{aa}$ being different
from $k_{bb}$, would result in favoring the attachment of say, an
$S$ to an $S$ over the attachment of an $R$ to an $R$, and this
feature would show clearly up in the relative abundances.

Another positive feature of the model is the robustness of the fits
with respect to differing initial imbalances of the enantiomers.
That is, for a given chemical model (including catalyst, if any) the
values of the fitted rates do not depend on the initial enantiomeric
excesses of the monomers. If our rate constants are viewed as
effective, that is, implicitly involving the different chemical
properties of the racemic and enantiomorphous crystallite phases,
then this feature suggests that the packing arrangements and
reaction pathways in the solid-state do not depend (or only weakly)
on the magnitude of these imbalances.

The Pearson product-moment correlation coefficient $r$ between
experimental and numerical data is greater for the copolymerization
model than for the binomial distribution, except for the
$C_{18}-thio-Glu$, which truly polymerizes randomly. The correlation
between calculated and experimental relative abundances is also
greater for the initially non-racemic situations, and the higher the
initially chiral enrichment of the mixture is, the better the
copolymerization model reproduces the chemical data. The results
obtained here lead us to affirm that the model systems considered
all undergo a non-random polymerization, as was asserted by the
authors of the experiments \cite{Lahav2002b,Lahav2003b}.

The model also qualitatively reproduces the behavior of the
enantiomeric excess $ee$, its increase with the length of the chains
and the enhancement of the $ee$ of the corresponding initial mixture
of monomers. All this, in spite of the complexity of the factors
that affect the reactivity within the experimental two-phase system,
i.e., the microscopic crystallite packing arrangements and the
possible reaction pathways within these 2D crystallites. In
conclusion then, we may therefore assert that our simple scheme does
provide an accurate course-grained description of the
lattice-controlled polymerization reported in Ref
\cite{Lahav2002b,Lahav2003b}.

\section*{Acknowledgements}
We are grateful to Meir Lahav for providing us with the experimental
data and for many helpful discussions and correspondence. CB has a
Calvo-Rod\'{e}s predoctoral scholarship from the Instituto Nacional
de T\'{e}cnica Aeroespacial (INTA) and the research of DH is
supported in part by the grant AYA2009-13920-C02-01 from the
Ministerio de Ciencia e Innovaci\'{o}n (Spain) and forms part of the
COST Action CM0703 ``Systems Chemistry".

%\footnotesize{
%\bibliography{copoly} %your .bib file
%\bibliographystyle{plain} %the RSC's .bst file

\bibliographystyle{cj}   % (uses file "plain.bst")
\bibliography{copoly}

\end{document}